\documentclass{article}
\usepackage{spconf,amsmath,graphicx}
\usepackage{amssymb}
\usepackage{enumitem}
\usepackage{hyperref}
\usepackage{amsmath}
\usepackage{comment}
\usepackage{amsmath,amssymb,epsfig}
\usepackage{bm}
\usepackage{verbatim}
\usepackage{mathabx}
\usepackage{setspace}
\usepackage{subfig}
\usepackage[x11names]{xcolor}
\usepackage[belowskip=2pt,aboveskip=10pt]{caption}

\ninept
\newcommand{\htt}{\ensuremath{\widetilde{h}}}
\newcommand{\st}{\ensuremath{\widetilde{s}}}
\newcommand{\xt}{\ensuremath{\widetilde{x}}}
\newcommand{\nt}{\ensuremath{\widetilde{n}}}
\newcommand{\Gt}{\ensuremath{\widetilde{G}}}
\newcommand{\Ht}{\ensuremath{\widetilde{H}}}
\newcommand{\St}{\ensuremath{\widetilde{S}}}
\newcommand{\Xt}{\ensuremath{\widetilde{X}}}

\newcommand{\alphat}{\ensuremath{\widetilde{\alpha}}}
\newcommand{\cW}{\ensuremath{\mathcal{W}}}
\newcommand{\cB}{\ensuremath{\mathcal{B}}}

\newcommand{\bM}{\ensuremath{\boldsymbol{M}}}

\newcommand{\bx}{\ensuremath{\boldsymbol{x}}}
\newcommand{\bg}{\ensuremath{\boldsymbol{g}}}
\newcommand{\bh}{\ensuremath{\boldsymbol{h}}}
\newcommand{\bn}{\ensuremath{\boldsymbol{n}}}
\newcommand{\bs}{\ensuremath{\boldsymbol{s}}}
\newcommand{\balpha}{\ensuremath{\boldsymbol{\alpha}}}
\newcommand{\bTheta}{\ensuremath{\boldsymbol{\Theta}}}
\newcommand{\bPhi}{\ensuremath{\boldsymbol{\Phi}}}
\newcommand{\bPsi}{\ensuremath{\boldsymbol{\Psi}}}
\newcommand{\bUpsilon}{\ensuremath{\boldsymbol{\Upsilon}}}
\newcommand{\bA}{\ensuremath{\boldsymbol{A}}}
\newcommand{\bI}{\ensuremath{\boldsymbol{I}}}
\newcommand{\bJ}{\ensuremath{\boldsymbol{J}}}
\newcommand{\bN}{\ensuremath{\boldsymbol{N}}}
\newcommand{\bT}{\ensuremath{\boldsymbol{T}}}
\newcommand{\bU}{\ensuremath{\boldsymbol{U}}}
\newcommand{\bzero}{\ensuremath{\boldsymbol{0}}}
\newcommand{\diag}{\ensuremath{\text{diag}}}
\newcommand{\bcH}{\ensuremath{\boldsymbol{\mathcal{H}}}}

\newif\ifproofread

\newcommand{\changemarker}[1]{%
	\ifproofread
	\textcolor{red}{#1}%
	\else
	#1%
	\fi
}

    \title{Multiresolution time-of-arrival estimation from multiband 
       radio channel measurements\footnotemark}

    \name{Tarik Kazaz, Raj Thilak Rajan, Gerard J. M. Janssen and Alle-Jan van der Veen
    \thanks{This research was supported in part by NWO-STW under 
    	contract 13970 (``SuperGPS'').}	}
    \address{Faculty of EEMCS, Delft University of Technology, Delft, The Netherlands}

\begin{document}
\proofreadfalse
\maketitle


\begin{abstract}
\noindent Achieving high resolution time-of-arrival (TOA)
    estimation in multipath propagation scenarios from bandlimited
    observations of communication signals is challenging because the
    multipath channel impulse response (CIR) is not bandlimited.
    Modeling the CIR as a sparse sequence of Diracs, TOA estimation becomes a problem of parametric spectral
    inference from observed bandlimited signals.  
    To increase resolution without arriving at unrealistic
    sampling rates, we consider multiband sampling approach, and propose a practical multibranch receiver for the acquisition. The resulting data model exhibits
    multiple shift invariance structures, and we propose a corresponding
    multiresolution TOA estimation algorithm based on the ESPRIT
    algorithm. 
    The performance of the algorithm is compared against the derived
    Cram\'er Rao Lower Bound, using simulations with standardized
    ultra-wideband (UWB) channel models.  We show that the proposed
    approach provides high resolution estimates while reducing spectral
    occupancy and sampling costs compared to traditional UWB approaches.

\end{abstract}

\section{Introduction}

    Time-of-arrival (TOA) estimation usually starts with the estimation
    of the underlying multipath communication channel.
    As the channel frequency response (CFR) is not bandlimited while we
    can only probe the channel with bandlimited signals, modeling
    assumptions are required.
    Traditionally, the channel impulse response (CIR) is modeled as an FIR
    filter of limited time duration, and the resulting time resolution
    for TOA estimation is inversely proportional to the sampling rate,
    i.e., to the bandwidth of the probing signal.  This motivates the
    use of ultra-wideband (UWB) systems 
    \cite{gezici2005localization, witrisal2009noncoherent}, but at 
    the cost of large spectrum occupancy,
    high sampling, and high computational requirements at the receiver.

    High resolution techniques therefore refine the channel model by
    considering a parametric model consisting of a small number of
    attenuated and delayed Diracs.  Under this assumption,
    theoretically we only need to take an equally small number of
    samples in the frequency domain.  The main challenge is to devise
    practical and robust schemes for implementing this.

    In the past, many delay estimation algorithms have been proposed, and they can be
    classified into methods based on (i) subspace estimation
    \cite{van1998joint, li2004super, pourkhaatoun2014high}, (ii) finite
    rate of innovation \cite{6596614, vetterli2002sampling,
    maravic2005sampling}, and (iii) compressed sampling signal
    reconstruction \cite{cohen2014channel, mishali2009blind, mishali2011xampling, zhang2009compressed,
    gedalyahu2010time, gedalyahu2011multichannel}.  Some of these methods are
    not quite robust to noise, while other methods require a separate
    receiver chain for each multipath component, which may not be practical.

    To improve resolution, a large frequency band (aperture) must be
    covered, while to limit sampling rates, the total band should not be
    densely sampled.  This motivates the use of multiband acquisition
    systems,  for e.g., \cite{wagner2012compressed} proposes estimation from
    a set of ``dispersed'' Fourier coefficients.  Other methods include for e.g.,
    bandpass sampling, multicoset sampling and modulated wideband
    converter (MWC) \cite{eldar2015sampling}, where the implementation at
    the analog front-end is not straight forward.

    In this paper, we aim at a limited complexity high resolution TOA
    estimation algorithm and consider coherent multiband acquisition.
    In a multichannel receiver, each receiver chain will coherently
    sample one of the sub-bands, which can be implemented with
    off-the-shelf radio frequency (RF) components.  
    By stacking the observations into Hankel matrices, the resulting
    data model has precisely the form of Multiple
    Invariance ESPRIT \cite{miesprit1992}, so that the related algorithms
    are applicable, in particular, the Multiresolution ESPRIT algorithm
    \cite{lemma1999multiresolution}, which was aimed at carrier
    frequency estimation.

    Similar to \cite{lemma1999multiresolution}, we propose an algorithm
    where the invariance structure of a single sub-band will provide
    coarse parameter estimates, while the the invariance structure of
    the lowest against the highest frequency sub-band will provide
    high-resolution, but phase wrapped, estimates. The wrapping is
    resolved using the coarse estimates.

    The resulting algorithm is benchmarked through simulations, by
    comparing its performance with the Cram\'er Rao Lower Bound (CRLB).
    The results show that the proposed approach provides high resolution
    estimates while reducing spectral occupancy and sampling costs
    compared to classical UWB approaches, paving the way for cognitive
    radio ranging systems.

\section{Problem Formulation and Data Model}
\label{sec:prob_form}

\noindent\textbf{Channel model:}
    We consider a channel model which is appropriate for modeling the
    multipath propagation of wideband and UWB signals. The multipath
    channel with $K$ propagation paths is defined by a continuous-time
    impulse response $\htt(t)$ and its continuous time frequency transform (CTFT)
    $\Ht(\Omega)$ as
\begin{equation}
\label{eq:chan_imp}
	\htt(t) = \sum_{k=1}^{K} \alphat_k \delta(t-\tau_k) 
	\quad \text{and} \quad 
	\Ht(\Omega) = \sum_{k=1}^{K} \alphat_k e^{-j\Omega\tau_k},
\end{equation}
\noindent 
    where we use ``tilde'' to represent signals at RF frequencies,
    $\alphat_k \in \mathbb{R}$ and $\tau_k \in \mathbb{R}_+$ represent
    the gain and time-delay of the $k$th resolvable path
    \cite{cramer2002evaluation}.  This model neglects the effects of
    frequency dependent distortions \cite{molisch2009ultra}.  However,
    for the purpose of our analysis, it provides sufficient
    characterization of the radio signal propagation.

\begin{figure}[t]
    \centering
    \subfloat[]{%
	    \includegraphics[trim=1 1 0 1,clip, width=6.6cm]{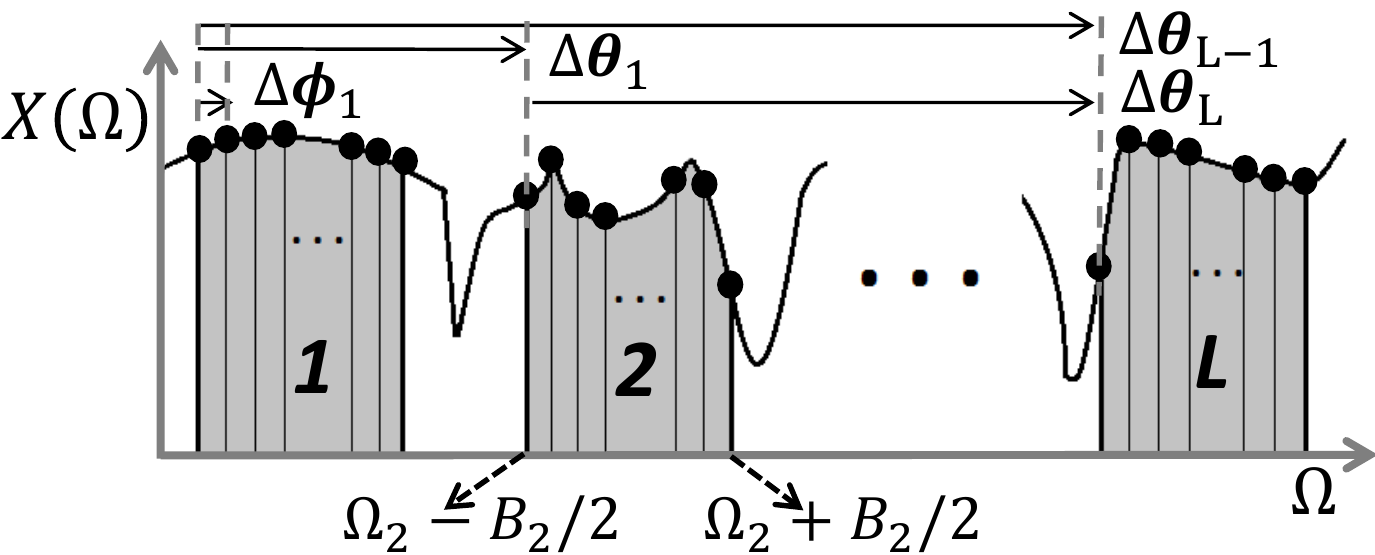}%
	    \label{fig:ch:res}%
    }\qquad
    \subfloat[]{%
	    \includegraphics[trim=0 1 1 2,clip,width=6.6cm]{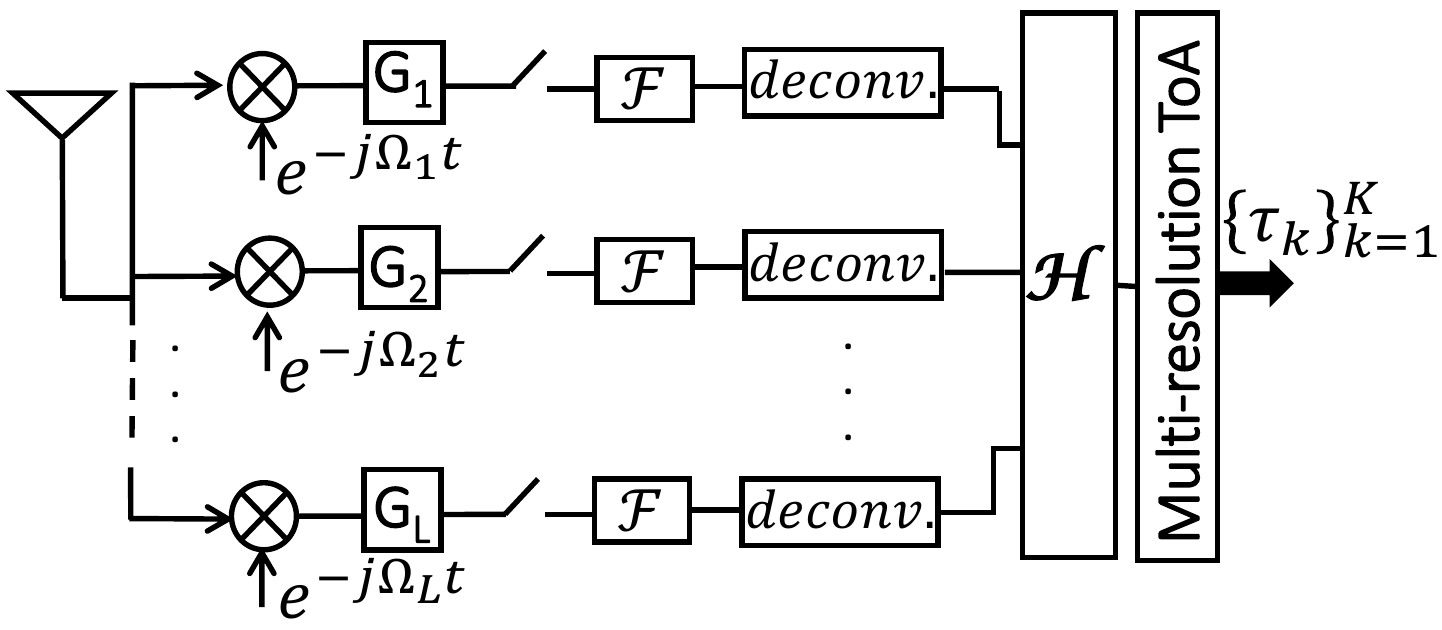}%
	    \label{fig:ch:rec}%
    }\qquad
\caption{(a) The multiband channel frequency response, and (b) a
    multibranch receiver with $L$ RF chains.}
    \label{fig:ch}
\end{figure}

\noindent\textbf{Continuous-time signal model:} 
    The objective in the paper is to estimate the $2K$ channel
    parameters by probing the channel by a wideband training signal
    $\st(t)$.  
    Assume that
    $\st(t)$ covers (at least) $L$ separate bands $\cW_i = [\Omega_i -
    \frac{1}{2} B_i, \Omega_i + \frac{1}{2} B_i]$, $i = 1, \cdots, L$,
    where $\Omega_i$ is the center frequency and $B_i$ is the bandwidth
    of the $i$th sub-band.
    The CTFT of $\st(t)$ is 
\begin{equation}
\label{eq:mb_sig}
   \St(\Omega)  =
   \begin{cases}
   \St_i(\Omega), & \Omega \in \cW_i, \; i = 1, \cdots, L
   \\
   \mbox{arbitrary}, & \text{otherwise}\,.
   \end{cases}
\end{equation}
   The received signal is $\xt(t) = \st(t) \ast \htt(t) + \nt(t)$, where
   $\nt(t)$ represents additive white Gaussian noise.
   The corresponding CTFT is (cf.\ Fig.\ \ref{fig:ch:res})
\begin{equation}
\label{eq:mb_rx}
   \widetilde{X}(\Omega)  =
   \widetilde{H}(\Omega)\widetilde{S}(\Omega) + \widetilde{N}(\Omega)
   \,.
\end{equation}
\noindent 
    Now, consider a multibranch receiver having $L$ RF chains and complex
    sampling ADCs as shown in Fig.\ \ref{fig:ch:rec}.  The $i$th RF chain
    bandlimits $\Xt(\Omega)$ to $\Omega \in \cW_i$, and
    performs complex downconversion to baseband, possibly followed by
    additional lowpass filtering before sampling at Nyquist.  
    We model this by an equivalent lowpass filter $G_i(\Omega)$ with
    passband $\cB_i = [-\frac{1}{2} B_i, \frac{1}{2} B_i]$.
    The CTFT of the signal $x_i(t) \in \mathbb{C}$ received in the $i$th
    branch at baseband is thus given by
\begin{equation}
\label{eq:mb_rx_bs}
   X_i(\Omega)  =
   \begin{cases}
   G_i(\Omega) H_i(\Omega) S_i(\Omega) + N_i(\Omega), & \Omega \in \cB_i\\
   0, & \text{otherwise}
   \end{cases}
\end{equation}
    where $N_i(\Omega)$ is bandlimited white Gaussian noise and
    $\{H_i(\Omega)$, $S_i(\Omega), G_i(\Omega)\}$ are the complex
    baseband equivalents of $\{\Ht(\Omega)$, $\St(\Omega), \Gt(\Omega)\}$.  In particular, $H_i(\Omega) =
    \Ht(\Omega+\Omega_i)$.

\noindent \textbf{Discrete-time data model:} 
    Assume for theoretical purposes, that $x_i(t)$ has a finite duration
    $T$, and is zero (or periodic) outside this interval.\footnote{In
    more general cases, a small bias will occur in the subsequent
    derivation.} 
    For simplicity of exposition, we will consider
    that the bandwidths of the signals and the sampling periods in all receiver branches are the same, that is $B_i=B$ and $T_{s,i}=T_s$ for all $i \in [1, L]$.  
    We sample $x_i(t)$ with period $T_s$, and take $N$ samples on the
    nonzero interval such that $T = N T_s$.  Let $\Omega_s = 2\pi/T_s$,
    then the $N$-point DFT of $x_i(t)$ is given by\footnote{A factor
    $1/T_s$ is absorbed in $S_i[n]$.}
\begin{equation}
\label{eq:dft_dtft}
   X_i[n] = G_i[n] H_i[n] S_i[n] + N_i[n],\quad n=0, \cdots, N-1
\end{equation}
    with, in particular,
\[
    H_i[n] = H_i\left(\frac{n}{N} \Omega_s\right) 
    = \Ht(\frac{n}{N} \Omega_s + \Omega_i) 
    \,.
\]
    Inserting the channel model (\ref{eq:chan_imp}) gives
\begin{equation}
\label{eq:chmodel}
    H_i[n] = \sum_{k=1}^{K} 
	\alphat_k e^{-j \Omega_i \tau_k} e^{-j n\Omega_t \tau_k}
\end{equation}
    where $\Omega_t = \frac{1}{N}\Omega_s = \frac{2\pi}{T}$.

    Let us stack the $N$ samples of $X_i[n]$ into a vector $\bx_i$, and
    likewise for $\bg_i$, $\bh_i$, $\bs_i$, and $\bn_i$.
    The data model (\ref{eq:dft_dtft}) then becomes
\begin{equation}
\label{eq:bx}
    \bx_i = \bh_i \odot \bg_i\odot\bs_i + \bn_i \,.
\end{equation}
    where $\odot$ denotes a pointwise multiplication. The channel model
    (\ref{eq:chmodel}) can be written as
\begin{equation}
\label{eq:chanmodel_matrix}
    \bh_i = \bM \bTheta_i \balpha \,,
\end{equation}
    where $\bM$ is the $N\times K$ Vandermonde matrix
\begin{equation}
\label{eq:f_matrix}
   \bM = 
   \begin{bmatrix}
       1 & 1 & \cdots & 1 \\
       \Phi_1 & \Phi_2 & \cdots & \Phi_K\\
       \vdots & \vdots & \ddots & \vdots \\ 
       \Phi_1^{{N-1}} & \Phi_2^{N-1} & \cdots & \Phi_K^{N-1}
   \end{bmatrix}
   \,,\quad
\end{equation}
   and $ \Phi_k = e^{-j\phi_k}$, where $\phi_k = \Omega_t \tau_k$. Likewise,
\begin{equation}
\label{eq:bTheta}
   \bTheta_i = 
   \begin{bmatrix}
    \Theta_{i,1} && \bzero\\
    &\ddots & \\
    \bzero && \Theta_{i,K}
   \end{bmatrix}
   \,,\qquad
   \balpha = \begin{bmatrix}
       \alphat_{1} \\ \vdots \\ \alphat_{K}
    \end{bmatrix}
\end{equation}
    and $\Theta_{i,k} = e^{-j\theta_{i,k}}$,
    where $\theta_{i,k} = \Omega_i\tau_k$.

   Next, we apply deconvolution to the data vector (\ref{eq:bx}). Assume 
   that no entry of $\bg_i$ and $\bs_i$ is zero or close to
   zero.\footnote{
   If there are zero entries, then we need to select a subvector of
   consecutive nonzero entries, and similar results will hold.}
   As the entries of these vectors are known from training and filter
   design/calibration, the deconvolution step to estimate the DFT channel
   coefficients is written as
\begin{equation}
\label{eq:deco}
   \bh_i = \{ \diag (\bg_i \odot \bs_i)\}^{-1} \bx_i,
\end{equation}
   which satisfies the model
\begin{equation}
\label{eq:model}
   \bh_i =  \bM \bTheta_i \alpha + \bn'_i,
\end{equation}
   where $\bn'_i$ is a zero mean circular symmetric complex Gaussian
   distributed noise vector.  It is common that the power spectral
   densities of the signal or the filters are not perfectly flat.  In
   that case, the noise vector is not white, but the coloring is known and
   can be taken into account.

\section{Multiresolution Delay Estimation} 
   Our next objective is to estimate the $K$ time-delays $\{{\tau_k}\}_{k=1}^K$.
   We begin with an algorithm for estimating these time-delays using a
   single frequency band, and later extend it for the multiple bands.
   The single band algorithm is in fact classical (cf.\ \cite{van1998joint, qian2016enhanced, steinwandt2017generalized} 
   and earlier references).

\subsection{Single band estimation algorithm} 
\label{sc:sb_estim} 

   From a single vector $\bh_i$, we construct a Hankel matrix of size
   $P\times Q$ as
\begin{equation}
\label{eq:hankel}
   \bcH_i = \begin{bmatrix}
      H_i[0] & H_i[1] & \cdots & H_i[Q] \\
      H_i[1] & H_i[2] & \cdots &   H_i[Q+1] \\
      \vdots &  \vdots   &   \ddots     & \vdots  \\
      H_i[P-1] & H_i[P] & \cdots & H_i[N-1]
    \end{bmatrix} \,.
\end{equation}
   Here, $P = N-Q-1$, and we require $P>K$ and $Q\ge K$.
   From (\ref{eq:model}), and using the shift invariance of the Vandermonde
   matrix (\ref{eq:f_matrix}), the Hankel matrix satisfies
\begin{equation} 
\label{eq:hen_mat}
   \bcH_i = \bM' \bTheta_i \bA + \bN_i \,,
\end{equation} 
   where $\bM'$ is an $P\times K$ submatrix of $\bM$, and
   $\boldsymbol{N}_i$ is a noise matrix.  Furthermore, 
\[
   \bA = [\balpha,\, \bPhi \balpha,\, \bPhi^2 \balpha, \cdots, \bPhi^{Q-1}
   \balpha] \,
\]
   where $\mathbf{\Phi} = \text{diag}([\phi_1 \cdots \phi_K])$.
   
   Since (\ref{eq:hen_mat}) resembles the data
   model of the classical ESPRIT algorithm, $\bPhi$ can be
   estimated by exploiting the low-rank approximation of the Hankel
   matrix and its shift-invariance properties. From $\bPhi$, the parameters
   $\tau_k$ immediately follow.

   In particular, let $\bU$ be a $K$-dimensional orthonormal basis for
   the column span of $\bcH_i$, obtained using the singular value
   decomposition, then we can write $\bM' = \bU \bT$, where $\bT$ is a
   $K\times K$ nonsingular matrix.
   Next, let us define selection matrices
   \begin{equation}
	   \label{eq:sel_mat}
	   \bJ_{1}^{(1)}  = [\bI_{P-r} \quad \bzero_{P-r,r} ], 
	   \qquad 
	   \bJ_{2}^{(1)} = [\bzero_{P-r,r} \quad \bI_{P-r}], 
   \end{equation} 
   
   \noindent where $\bI_{P-r}$ is identity matrix of size $(P-r) \times (P-r)$ and $\mathbf{0}_{P-r,r}$ is
   a zero matrix of size $(P-r) \times r$. 
    For $r=1$, $\bU_1 = \bJ_{1}^{(1)} \bU$ and $\bU_2 = \bJ_{2}^{(1)} \bU$
    are submatrices of $\bU$ obtained by dropping its first and, respectively, last
    row.  In view of the shift invariance
    structure of $\bM'$, we have
\begin{equation}
\label{eq:fac_sub}
   \bU_1 = \bM_1' \bT^{-1}\,,\qquad
   \bU_2 = \bM_1' \bPhi \bT^{-1}
\end{equation} 
   where $\bM_1'= \bJ_{1}^{(1)} \bM'$. Finally, we form 
   the matrix $\bPsi = \bU_1^\dagger \bU_2$ where $\dagger$ denotes
   pseudo-inverse. $\bPhi$ can then be estimated directly
   from the eigenvalue decomposition of $\bPsi$, 
   as it satisfies the model
\begin{equation}
\label{eq:eig}
   \bPsi = \bT\bPhi \bT^{-1}  \,.
\end{equation} 
    In other words, let $\hat{\lambda}_k$ be an estimate of the $k$th
    eigenvalue of $\bPsi$, then the corresponding time delay estimate is
    ${\tau}_k = \text{arg}\{\lambda_k\}/\Omega_t$.  Since $\Omega_t \tau_k
    < 2\pi$ because $\tau_k < T$, there is no phase wrapping issue here.
    Note that for TOA estimation, we are mostly interested in retrieving 
    the smallest $\tau_k$ as it belongs to the line-of-sight propagation, i.e., true distance. 
  
\subsection{Multiresolution estimation algorithm} 
\label{sc:mb_estim}
   The aforementioned algorithm used data from a single sub-band and has a limited
   resolution, since it is based on the shift of one row in the Hankel
   matrix $\bcH_i$, which results in only a small phase shift $\Omega_t
   \tau_k$.  Note that the sampling rate does not play a role in
   $\Omega_t$, only the total signal duration $T$. \changemarker{Thus, oversampling would increase the signal-to-noise ratio (SNR) but not the resolution.}

   The matrix $\bM$ is also invariant for shifts over multiple rows, and
   therefore, if $N$ is sufficiently large, 
   then we can increase the resolution by considering shifts of multiple rows of
   $\bcH_i$. 
   Indeed, a shift of $r$ rows using shift matrices $\bJ_{1}^{(r)}, \bJ_{2}^{(r)}$
   (or by interleaving rows of $\boldsymbol{\mathcal{H}}_i$
   \cite{maravic2005sampling})
   leads to an estimate of $\bPhi^r$.
   Unfortunately, phase shifts have an ambiguity of multiples of $2\pi$,
   so that approaches for increasing the resolution introduce ambiguity
   in the estimates for the $\tau_k$. If $T$ is not very large, this
   approach is limited.

   Here, we are interested in an algorithm for high resolution and
   unambiguous estimation of the $\tau_k$ from multiband channel estimates
   $\bh_i$, where $i = 1, \cdots, L$.  For simplicity of exposition,
   we will consider for the moment only two
   bands (i.e.,  $i = 1, 2$), with central frequencies $\Omega_1$ and
   $\Omega_2$.  Following the procedure described in Section
   \ref{sc:sb_estim}, we form the Hankel matrices
   $\bcH_i$ defined in (\ref{eq:hankel}) and stack them in a
   matrix 
\begin{equation}
\label{eq:mb_mat} 
    \nonumber 
    \bcH =
    \begin{bmatrix} \boldsymbol{\mathcal{H}}_1 \\
	\boldsymbol{\mathcal{H}}_2 
    \end{bmatrix}.
\end{equation} 
   The matrix $\boldsymbol{\mathcal{H}}$ has the model 
\begin{equation} 
\label{eq:mb_fac} 
    \bcH =
    \begin{bmatrix} 
       \bM' \\
       \bM' \bTheta
   \end{bmatrix}
   \bTheta_1\bA  + \bN\,,
\end{equation}
   where $\bTheta = \bTheta_2\bTheta_1^{-1}$,
   $\boldsymbol{\Theta}_1$ and $\boldsymbol{\Theta}_2$ are given in
   (\ref{eq:bTheta}), and $\bN$ is formed by stacking
   $\bN_1$ on top of $\bN_2$.  Note that
   $\bcH$ has a double shift invariance structure
   introduced by the phase shifts of the $\tau_k$ on the 
   (i) sampling frequency within a single band, $\phi_k$, and (ii) carrier
   frequency difference between two bands,
   $\theta_k=\theta_{2,k}-\theta_{1,k}$, as
   shown in Fig.\  \ref{fig:ch:res} \cite{6489262}.  In general, the
   carrier frequency difference is much higher than the sampling
   frequency, and therefore, $\theta_k\gg\phi_k$ for $k = 1,\cdots,
   K$.  The estimation of the $\tau_k$ from
   $\bTheta$ will result in high resolution but
   ambiguous estimates, due to unknown multiples of $2\pi$ in the phases.
   However, we can use the idea of multiresolution
   parameter estimation \cite{lemma1999multiresolution}
   to develop the algorithm for high resolution
   estimation of the $\tau_k$ without ambiguity by combining
   coarse and fine estimates obtained from $\bPhi$ and
   $\bTheta$, respectively.

   We follow a similar approach as in the previous section. Let $\bU$ be
   an orthonormal basis for the column span of $\bcH$, obtained using 
   a truncated SVD.
   Define the selection matrices 
\begin{equation} 
\label{eq:sel_mat2}
   \begin{aligned} \centering 
       \bJ_{\Phi1}^{(r)} = \bI_2 \otimes [\bI_{P-r} \quad \bzero_{P-r,r}], 
   & \qquad
       \bJ_{\Theta1} = [1 \quad 0] \otimes \bI_{P}, 
   \\
       \bJ_{\Phi2}^{(r)} = \bI_2 \otimes [\bzero_{P-r,r} \quad \bI_{P-r}], 
   & \qquad 
       \bJ_{\Theta2} = [0 \quad 1] \otimes \bI_{P}.  
   \end{aligned} 
\end{equation}
   To estimate $\bPhi$, we set $r=1$ and take submatrices
   consisting of the first and, respectively, the last row of each block
   matrix stacked in $\bU$, that is
   $\bU_{\Phi1}=\bJ_{\Phi 1}^{(1)}\bU$ 
   and $\bU_{\Phi 2}=\bJ_{\Phi2}^{(1)}\bU$.  The estimation of
   $\bTheta$ is based on the first and, respectively,
   second block matrix present in $\bU$, that is
   $\bU_{\Theta 1}=\bJ_{\Theta1}\bU$ and
   $\bU_{\Theta2}=\bJ_{\Theta 2}\bU$. 
   The selected matrices have the following models:
\begin{equation} 
   \begin{aligned} 
   \label{eq:fac_sel}
       \bU_{\Phi 1} &= 
          \begin{bmatrix} \bM'' \\ \bM'' \bTheta \end{bmatrix}
       \bTheta_{1} \bT^{-1}, 
   & \qquad
       \bU_{\Theta1} &= \bM' \bT^{-1}, 
   \\
       \bU_{\Phi 2} &= 
       \begin{bmatrix} \bM'' \\ \bM'' \bTheta \end{bmatrix}
       \bPhi\bTheta_1 \bT^{-1}, 
   & \qquad
       \bU_{\Theta2} &= \bM'\bTheta\bT^{-1},
   \end{aligned}
\end{equation} 
   where $\bM'' = \bJ_{1}^{(1)} \bM'$ and $\bJ_{1}^{(1)}$ is given in (\ref{eq:sel_mat}).  
   The Least Squares approximate solutions to the set of
   equations in (\ref{eq:fac_sel}) satisfy a model of the form
\begin{equation} 
   \begin{aligned} 
   \label{eq:mul_mod} 
       \bPsi &:= \bU_{\Phi1}^{\dagger} \bU_{\Phi2} = \bT\bPhi\bT^{-1}, 
       \\
       \bUpsilon &:= \bU_{\Theta1}^{\dagger} \bU_{\Theta2} = \bT\bTheta\bT^{-1}.
   \end{aligned} 
\end{equation} 

   \begin{figure*}[t!] \centering \subfloat[]{%
			\includegraphics[trim=2 2 1
			1,clip,width=6.0cm]{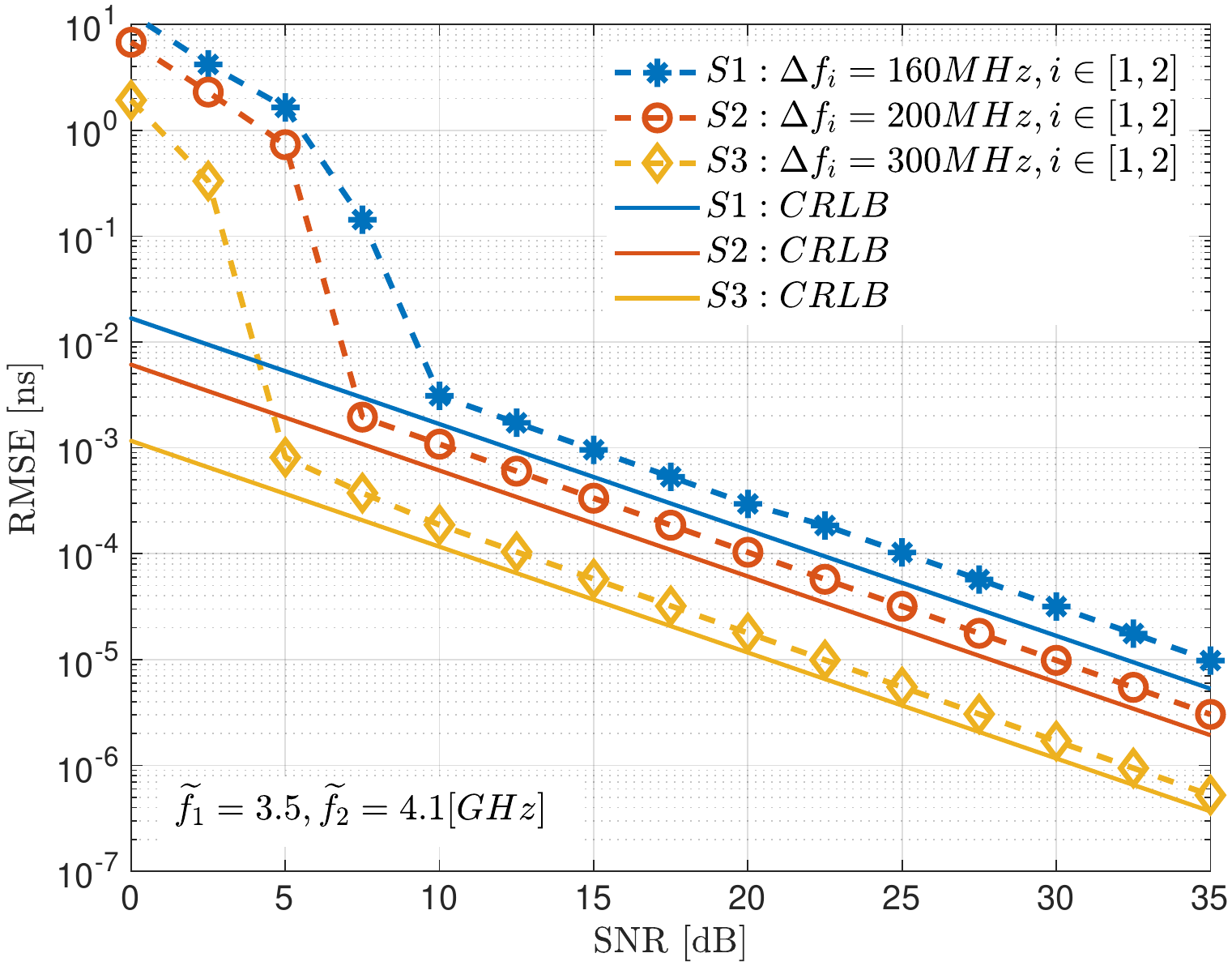}%
			\label{fig:per:bw}%
		} \subfloat[]{%
			\includegraphics[trim=2 2 1
			1,clip,width=6.0cm]{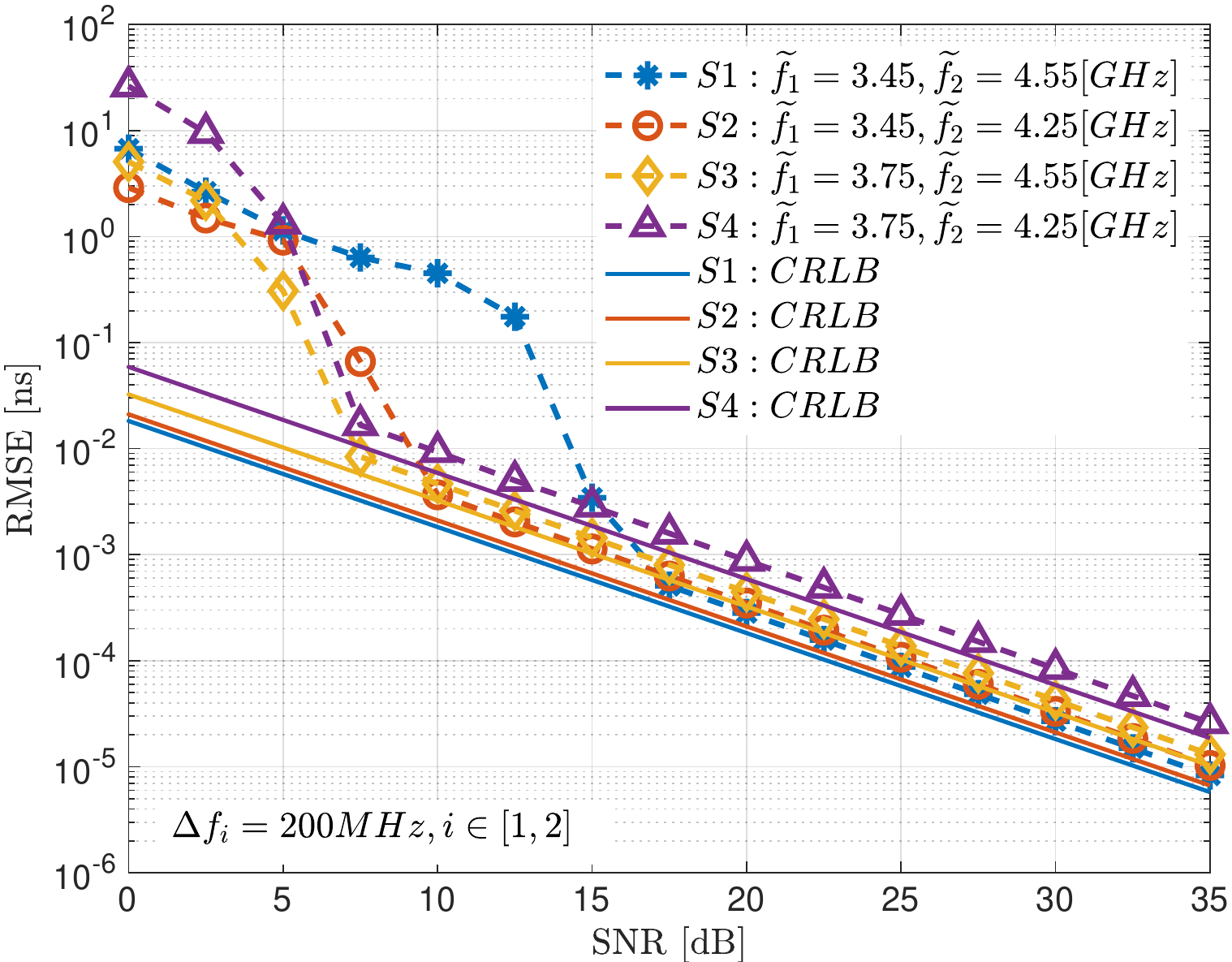}%
			\label{fig:per:pos}%
		} \subfloat[]{%
			\includegraphics[trim=2 2 1
			1,clip,width=6.0cm]{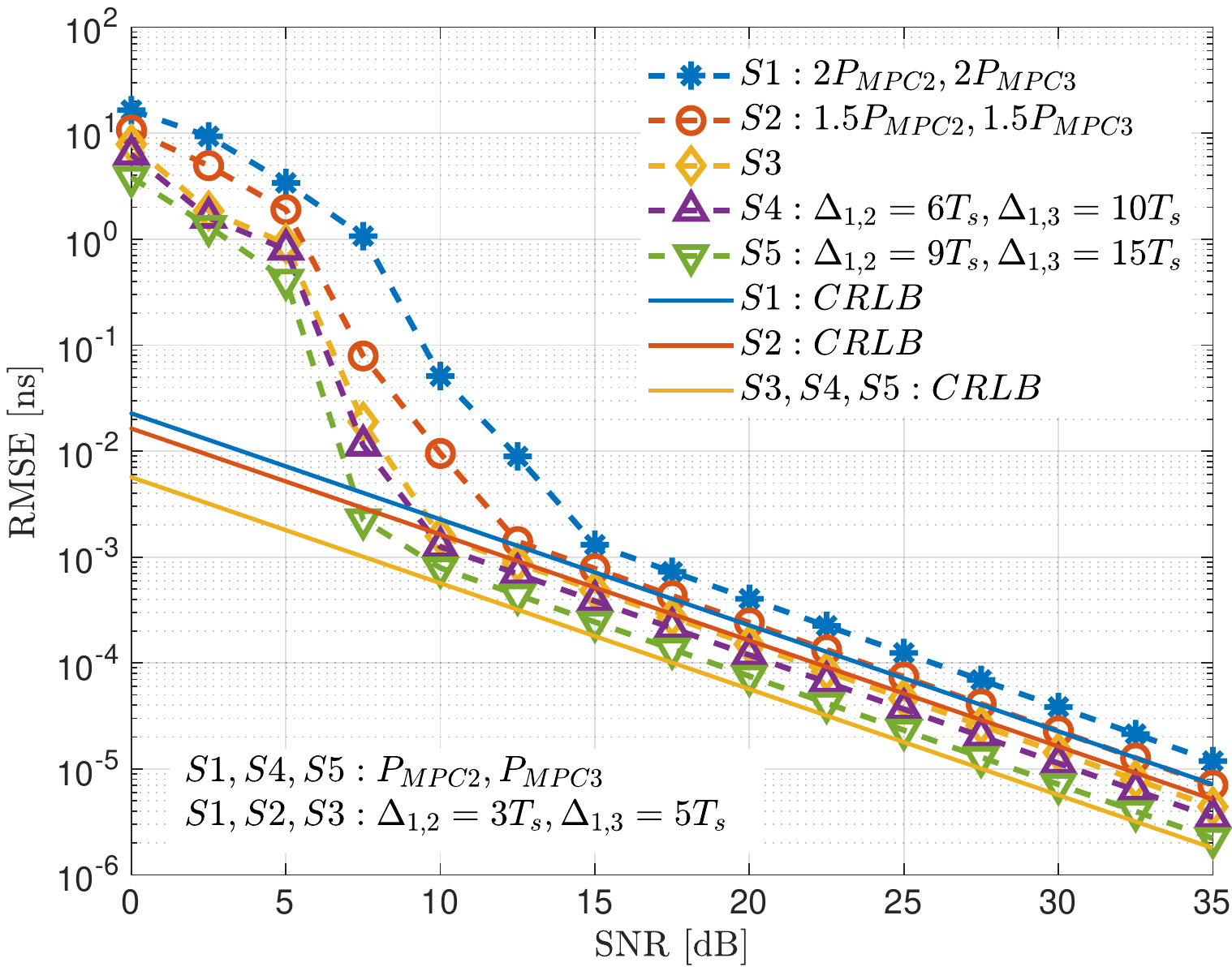}%
			\label{fig:per:pow_time}%
		} \caption{Root Mean Square Error (RMSE) of TOA estimates
			(${\tau}_1 $) for: (a) varying bandwidths, (b) varying
			band positions and (c) varying power and spacing of second and
			third MPC.}
		
		\label{fig:per} 
	\end{figure*}  

   Observe that $\bPsi$ and $\bUpsilon$ are jointly diagonalizable by the same matrix
   $\bT$.  If each submatrix in (\ref{eq:fac_sel}) has at least $K$
   rows, the joint diagonalization can be computed by means of QZ
   iterations or Jacobi iterations \cite{van1992azimuth,
   van1996analytical, chabriel2014joint}.  After $\bT$ has been
   determined, the parameters $\phi_k$ and $\theta_k$ for $k=1,
   \cdots, K$ are estimated.

   Based on the phase estimates, coarse and fine time-delays of the
   delays are computed as
\begin{equation} 
\label{eq:mul_est} 
    \tau_k = \Omega_t^{-1} \phi_k
	   = (\Omega_2-\Omega_1)^{-1} (\theta_k + 2 \pi n_k).
\end{equation}  

   The unknown number of cycles $n_k$ is determined as the best fitting
   integer that satisfies (\ref{eq:mul_est}), that is, 
\begin{equation}
\label{eq:mul_fit} 
   n_k =
   \text{round} \left\{
   \frac{1}{2\pi}
   \left(\Omega_t^{-1}(\Omega_2-\Omega_1)\phi_k-\theta_k\right)\right\}.
\end{equation} 
   If the estimation errors of the $\phi_k$ and the $\theta_k$ are
   comparable, then the $\tau_k$ estimate based on $\theta_k$ is
   $\Omega_t^{-1}(\Omega_2-\Omega_1)$ times more accurate and less
   affected by noise compared to the one based on $\phi_k$.  Therefore,
   the final estimate of $\phi_k$ is obtained based on $\theta_k$, or by
   optimal combining of coarse and fine estimates
   \cite{lemma1999multiresolution}.

   This technique can be extended to $L$ matrices. Alternatively, we only
   consider pairwise estimates, gradually increasing the resolution until
   we are able to reliably estimate the amount of $2\pi$ phase wraps for the
   largest shift ($\Omega_L - \Omega_1$).

\section{Results} 
\subsection{Cram\'er Rao Lower Bound (CRLB)} 
   We use the CRLB as a benchmark to study the performance of the algorithm derived
   in Section \ref{sc:mb_estim}.  The model (\ref{eq:model}) can be
   written as 
\begin{equation}
\label{eq:new_model}
   \mathbf{h}_i=\mathbf{B}_i \boldsymbol{\alpha} + \mathbf{n}_i,
\end{equation}
   where $\mathbf{B}_i = \mathbf{M} \boldsymbol{\Theta}_i =
   [\mathbf{m}_1\Theta_{i,1}, \; \mathbf{m}_2\Theta_{i,2}, \; \dots \;
   \mathbf{m}_K\Theta_{i,K}]$,  \changemarker{$\mathbf{m}_k$ for $k = 1, \cdots, K$ are the columns of $\mathbf{M}$} and $\mathbf{n}_i \sim
   \mathcal{C}\mathcal{N}(\mathbf{0}, \sigma^2 \mathbf{I}_N)$. The $\mathbf{B}_i$ is parameterized by $\boldsymbol{\tau} = [\tau_1,
   \cdots , \tau_K]^T$.  Under the assumption that the unknown multipath
   parameters $\boldsymbol{\alpha}$ and $\boldsymbol{\tau}$ are
   deterministic, the CRLB for estimating of $\boldsymbol{\tau}$,
   conditioned on complex path attenuations $\boldsymbol{\alpha}$, is
   given by \cite{stoica1989music}
\begin{equation} 
\label{eq:crlb} 
   \text{CRLB}(\boldsymbol{\tau})=
   \dfrac{\sigma^2}{2} \left\{ \Re \left[ \mathbf{D}^H
   \mathbf{P}_{\mathbf{B}_i}^\perp\mathbf{D} \odot
   \mathbf{R}_{\boldsymbol{\alpha}} \right]^{-1} \right\},
\end{equation} 
   where $\mathbf{D} = \left[ \dfrac{\partial
   \mathbf{b}_i(\tau_1)}{\partial \tau_1}, \dots , \dfrac{\partial
   \mathbf{b}_i(\tau_K)}{\partial \tau_K} \right]$,
   $\mathbf{b}_i(\tau_k) = \mathbf{m}_k\theta_{i,k}$ is the $k$th column
   of $\mathbf{B}_i$, $\mathbf{P}_{\mathbf{B}_i}^\perp = \mathbf{I}_N -
   \mathbf{B}_i(\mathbf{B}_i^H \mathbf{B}_i)^{-1}\mathbf{B}_i$ and
   $\mathbf{R}_{\boldsymbol{\alpha}} = E \{\boldsymbol{\alpha}
   \boldsymbol{\alpha}^H\}$.  It is straightforward to extend the CRLB
   for the multiband case by creating the overall data model in the form
   (\ref{eq:new_model}).

 \subsection{Simulations}
 We consider a standard outdoor UWB channel model to evaluate the
 performance of the proposed algorithm \cite{molisch2006comprehensive}.
 The channel (\ref{eq:chan_imp}) has eight dominant multipath components
 (MPCs).  The first MPC
 has $8$ times higher power in comparison to the second MPC.  The
 continuous time is modeled using a $3$~GHz grid, where the channel tap
 delays are spaced at $333.33$~ps.  In the simulations, we assume that
 the TOA is estimated using two bands with central frequencies at
 $\widetilde{f}_i = \Omega_i/(2\pi)$ and bandwidths $\Delta f_i =
 B_i/(2\pi)$ for $i \in [1,2]$.  We use Root Mean Square Error (RMSE)
 as a metric for evaluation, which is obtained over $10^4$ independent
 Monte Carlo runs.  These results are compared against the
 numerically computed CRLB (\ref{eq:crlb}).

   In Fig. \ref{fig:per:bw}, the RMSEs of the estimated TOAs (${\tau}_1$) for the first MPC are plotted against SNR for the frequency bands with bandwidths $[160, 200, 300]$ MHz. The proposed algorithm  asymptotically achieves the CRLB, for increasing SNR. As expected, for larger bandwidths the proposed algorithm is more robust to noise, and offers higher resolution. 
   
   In Fig.\ \ref{fig:per:pos}, the RMSEs of ${\tau}_1$ are plotted against SNR, for various positions of the $200$ MHz wide bands. It can be seen that by increasing the distance between two bands, i.e. frequency aperture, the resolution of the ${\tau}_1$ increases for high SNR. However, for low SNR the proposed algorithm has better performance in scenarios where the frequency aperture is lower which is a consequence of lower error for fine time-delay estimation.
   
   Fig.\ \ref{fig:per:pow_time} shows the RMSEs of the ${\tau}_1$ with respect to SNR for the following scenarios. Firstly, in scenarios S1 and S2, we consider the power of the second ($P_{MPC2}$) and third ($P_{MPC3}$) multipath components increased 2 or 1.5 times as compared to their value in S3, respectively. In scenarios S4 and S5 the distance between main and second ($\Delta_{1,2}$) and third ($\Delta_{1,3}$) MPC has been increased 2 or 3 times as compared to S3, respectively. As expected, the proposed algorithm is less robust to noise and has a lower resolution for scenarios where close MPCs have high power. It is seen, that the resolution of ${\tau}_1$ increases in scenarios where the main MPCs are more separated. 
   


\bibliographystyle{IEEEtran}
\newpage
\bibliography{refs}

\begin{thebibliography}{10}
\providecommand{\url}[1]{#1}
\csname url@samestyle\endcsname
\providecommand{\newblock}{\relax}
\providecommand{\bibinfo}[2]{#2}
\providecommand{\BIBentrySTDinterwordspacing}{\spaceskip=0pt\relax}
\providecommand{\BIBentryALTinterwordstretchfactor}{4}
\providecommand{\BIBentryALTinterwordspacing}{\spaceskip=\fontdimen2\font plus
\BIBentryALTinterwordstretchfactor\fontdimen3\font minus
  \fontdimen4\font\relax}
\providecommand{\BIBforeignlanguage}[2]{{%
\expandafter\ifx\csname l@#1\endcsname\relax
\typeout{** WARNING: IEEEtran.bst: No hyphenation pattern has been}%
\typeout{** loaded for the language `#1'. Using the pattern for}%
\typeout{** the default language instead.}%
\else
\language=\csname l@#1\endcsname
\fi
#2}}
\providecommand{\BIBdecl}{\relax}
\BIBdecl

\bibitem{gezici2005localization}
S.~Gezici \emph{et~al.}, ``Localization via ultra-wideband radios: a look at
  positioning aspects for future sensor networks,'' \emph{IEEE signal
  processing magazine}, vol.~22, no.~4, pp. 70--84, 2005.

\bibitem{witrisal2009noncoherent}
K.~Witrisal \emph{et~al.}, ``Noncoherent ultra-wideband systems,'' \emph{IEEE
  Signal Processing Magazine}, vol.~26, no.~4, 2009.

\bibitem{van1998joint}
A.-J. Van~der Veen, M.~C. Vanderveen, and A.~Paulraj, ``Joint angle and delay
  estimation using shift-invariance techniques,'' \emph{IEEE Transactions on
  Signal Processing}, vol.~46, no.~2, pp. 405--418, 1998.

\bibitem{li2004super}
X.~Li and K.~Pahlavan, ``Super-resolution {TOA} estimation with diversity for
  indoor geolocation,'' \emph{IEEE Transactions on Wireless Communications},
  vol.~3, no.~1, pp. 224--234, 2004.

\bibitem{pourkhaatoun2014high}
M.~Pourkhaatoun and S.~A. Zekavat, ``High-resolution low-complexity
  cognitive-radio-based multiband range estimation: {C}oncatenated spectrum vs.
  {F}usion-based,'' \emph{IEEE Systems Journal}, vol.~8, no.~1, pp. 83--92,
  2014.

\bibitem{6596614}
I.~Maravic, J.~Kusuma, and M.~Vetterli, ``Low-sampling rate {UWB} channel
  characterization and synchronization,'' \emph{Journal of Communications and
  Networks}, vol.~5, no.~4, pp. 319--327, Dec 2003.

\bibitem{vetterli2002sampling}
M.~Vetterli, P.~Marziliano, and T.~Blu, ``Sampling signals with finite rate of
  innovation,'' \emph{IEEE transactions on Signal Processing}, vol.~50, no.~6,
  pp. 1417--1428, 2002.

\bibitem{maravic2005sampling}
I.~Maravic and M.~Vetterli, ``Sampling and reconstruction of signals with
  finite rate of innovation in the presence of noise,'' \emph{IEEE Transactions
  on Signal Processing}, vol.~53, no.~8, pp. 2788--2805, 2005.

\bibitem{cohen2014channel}
K.~M. Cohen \emph{et~al.}, ``Channel estimation in {UWB} channels using
  compressed sensing,'' in \emph{Acoustics, Speech and Signal Processing
  (ICASSP), 2014 IEEE International Conference on}.\hskip 1em plus 0.5em minus
  0.4em\relax IEEE, 2014, pp. 1966--1970.

\bibitem{mishali2009blind}
M.~Mishali and Y.~C. Eldar, ``Blind multiband signal reconstruction:
  {Compressed} sensing for analog signals,'' \emph{IEEE Transactions on signal
  processing}, vol.~57, no.~3, pp. 993--1009, 2009.

\bibitem{mishali2011xampling}
M.~Mishali, Y.~C. Eldar, and A.~J. Elron, ``Xampling: {Signal} acquisition and
  processing in union of subspaces,'' \emph{IEEE Transactions on Signal
  Processing}, vol.~59, no.~10, pp. 4719--4734, 2011.

\bibitem{zhang2009compressed}
P.~Zhang \emph{et~al.}, ``A compressed sensing based ultra-wideband
  communication system,'' in \emph{Communications, 2009. ICC'09. IEEE
  International Conference on}.\hskip 1em plus 0.5em minus 0.4em\relax IEEE,
  2009, pp. 1--5.

\bibitem{gedalyahu2010time}
K.~Gedalyahu and Y.~C. Eldar, ``Time-delay estimation from low-rate samples:
  {A} union of subspaces approach,'' \emph{IEEE Transactions on Signal
  Processing}, vol.~58, no.~6, pp. 3017--3031, 2010.

\bibitem{gedalyahu2011multichannel}
K.~Gedalyahu, R.~Tur, and Y.~C. Eldar, ``Multichannel sampling of pulse streams
  at the rate of innovation,'' \emph{IEEE Transactions on Signal Processing},
  vol.~59, no.~4, pp. 1491--1504, 2011.

\bibitem{wagner2012compressed}
N.~Wagner, Y.~C. Eldar, and Z.~Friedman, ``Compressed beamforming in ultrasound
  imaging,'' \emph{IEEE Transactions on Signal Processing}, vol.~60, no.~9, pp.
  4643--4657, 2012.

\bibitem{eldar2015sampling}
Y.~C. Eldar, \emph{Sampling theory: {Beyond} bandlimited systems}.\hskip 1em
  plus 0.5em minus 0.4em\relax Cambridge University Press, 2015.

\bibitem{miesprit1992}
A.~L. Swindlehurst \emph{et~al.}, ``Multiple invariance {ESPRIT},'' \emph{IEEE
  Transactions on Signal Processing}, vol.~40, no.~4, pp. 867--881, 1992.

\bibitem{lemma1999multiresolution}
A.~N. Lemma, A.-J. Van~der Veen, and E.~F. Deprettere, ``Multiresolution
  {ESPRIT} algorithm,'' \emph{IEEE Transactions on signal processing}, vol.~47,
  no.~6, pp. 1722--1726, 1999.

\bibitem{cramer2002evaluation}
R.-M. Cramer, R.~A. Scholtz, and M.~Z. Win, ``Evaluation of an ultra-wide-band
  propagation channel,'' \emph{IEEE Transactions on Antennas and Propagation},
  vol.~50, no.~5, pp. 561--570, 2002.

\bibitem{molisch2009ultra}
A.~F. Molisch, ``Ultra-wide-band propagation channels,'' \emph{Proceedings of
  the IEEE}, vol.~97, no.~2, pp. 353--371, 2009.

\bibitem{qian2016enhanced}
C.~Qian \emph{et~al.}, ``Enhanced {PUMA} for direction-of-arrival estimation
  and its performance analysis,'' \emph{IEEE Transactions on Signal
  Processing}, vol.~64, no.~16, pp. 4127--4137, 2016.

\bibitem{steinwandt2017generalized}
J.~Steinwandt, F.~Roemer, and M.~Haardt, ``Generalized least squares for
  {ESPRIT}-type direction of arrival estimation,'' \emph{IEEE Signal Processing
  Letters}, vol.~24, no.~11, pp. 1681--1685, 2017.

\bibitem{6489262}
T.~Kazaz \emph{et~al.}, ``Joint ranging and clock synchronization for dense
  heterogeneous {IoT} networks,'' in \emph{2018 52nd Asilomar Conference on
  Signals, Systems, and Computers}, Oct 2018.

\bibitem{van1992azimuth}
A.-J. van~der Veen, P.~B. Ober, and E.~F. Deprettere, ``Azimuth and elevation
  computation in high resolution {DOA} estimation,'' \emph{IEEE Transactions on
  Signal Processing}, vol.~40, no.~7, pp. 1828--1832, 1992.

\bibitem{van1996analytical}
A.-J. Van Der~Veen and A.~Paulraj, ``An analytical constant modulus
  algorithm,'' \emph{IEEE Transactions on Signal Processing}, vol.~44, no.~5,
  pp. 1136--1155, 1996.

\bibitem{chabriel2014joint}
G.~Chabriel \emph{et~al.}, ``Joint matrices decompositions and blind source
  separation: A survey of methods, identification, and applications,''
  \emph{IEEE Signal Processing Magazine}, vol.~31, no.~3, pp. 34--43, 2014.

\bibitem{stoica1989music}
P.~Stoica and A.~Nehorai, ``Music, maximum likelihood, and cramer-rao bound,''
  \emph{IEEE Transactions on Acoustics, Speech, and Signal Processing},
  vol.~37, no.~5, pp. 720--741, 1989.

\bibitem{molisch2006comprehensive}
A.~F. Molisch \emph{et~al.}, ``A comprehensive standardized model for
  ultrawideband propagation channels,'' \emph{IEEE Transactions on Antennas and
  Propagation}, vol.~54, no.~11, pp. 3151--3166, 2006.

\end{thebibliography}

\end{document}